\documentclass[conference]{IEEEtran}
\IEEEoverridecommandlockouts
	\usepackage[english]{babel}
\usepackage{theorem}
\theoremheaderfont{\itshape\bfseries}
{\theorembodyfont{\itshape}
	
	\newtheorem{lemma}{\textbf{Lemma}}
	\newtheorem{definition}{\textbf{Definition}}
	\newtheorem{theorem}{\textbf{Theorem}}
	
	\newtheorem{remark}{\textbf{Remark}}
	
	\newtheorem{problem}{\textbf{Problem}}
}

\usepackage{graphicx}
\usepackage{epsfig}
\usepackage{newtxtext,newtxmath}
\usepackage{amsmath}
\usepackage{amsfonts}
\usepackage{mathrsfs}
\usepackage{xcolor}
\usepackage{graphicx}
\usepackage[ruled,vlined]{algorithm2e}
\usepackage{silence}
\usepackage{caption}
\usepackage[most]{tcolorbox}

\newcommand{\T}{^{\mbox{\tiny T}}}
\newcommand{\R}{\mathbb{R}}

\newcommand{\eps}{\varepsilon}
\let\leq\leqslant
\let\geq\geqslant

\newenvironment{proof}[1][Proof]%
{\par\noindent\textit{#1:\ }}%
{\hspace*{\fill} \rule{6pt}{6pt}}
\newenvironment{proof*}[1][Proof]%
{\par\noindent\textit{#1:\ }}{}

\DeclareMathOperator{\diag}{diag}

\DeclareMathOperator{\sat}{sat}
\DeclareMathOperator{\sgn}{sgn}

\usepackage{tikz}
\usetikzlibrary{shapes,calc,arrows,patterns,decorations.pathmorphing
	,decorations.markings} \usetikzlibrary{arrows.meta}

\newenvironment{system}[1]%
{\setlength{\arraycolsep}{0.5mm}\left\{ \; \begin{array}{#1}}%
	{\end{array} \right.}
\newenvironment{system*}[1]%
{\setlength{\arraycolsep}{0.5mm} \begin{array}{#1}}%
	{\end{array}}
\def\BibTeX{{\rm B\kern-.05em{\sc i\kern-.025em b}\kern-.08em
    T\kern-.1667em\lower.7ex\hbox{E}\kern-.125emX}}
\begin{document}

\title{Scale-free linear protocol design for global regulated state synchronization of discrete-time double-integrator multi-agent systems subject to actuator saturation
}

\author{\IEEEauthorblockN{Zhenwei Liu}
\IEEEauthorblockA{\textit{College of Information} \\
	\textit{Science and Engineering,} \\
\textit{Northeastern University}\\
Shenyang, China \\
liuzhenwei@ise.neu.edu.cn}
\and
\IEEEauthorblockN{Ali Saberi}
\IEEEauthorblockA{\textit{School of Electrical Engineering} \\
	\textit{and Computer
		Science,}\\
\textit{Washington State University}\\
Pullman, WA, USA \\
saberi@wsu.edu}
\and
\IEEEauthorblockN{Anton A. Stoorvogel}
\IEEEauthorblockA{\textit{Department of Electrical Engineering,} \\
	\textit{Mathematics and Computer Science,} \\
\textit{University of Twente}\\
Enschede, The Netherlands \\
A.A.Stoorvogel@utwente.nl}

}

\maketitle

\begin{abstract}
This paper studies global regulated state synchronization of discrete-time double-integrator multi-agent systems subject to actuator saturation by utilizing
localized information exchange. We propose a scale-free linear protocol that achieves global regulated state synchronization for any network with arbitrary number of agents and arbitrarily directed communication graph that has a path between each agent and exosystem which generates the reference trajectory. 
\end{abstract}


\section{Introduction}
The synchronization or consensus problem of multi-agent systems (MAS) has
drawn a lot of investigation in recent years, because of its developing applications in a few fields, like  automotive vehicle
control, satellites/robots formation, sensor networks, and so forth. The work can be seen, for example in the books
\cite{bai-arcak-wen,bullobook,kocarev-book, mesbahi-egerstedt,ren-book,
	saberi-stoorvogel-zhang-sannuti-book,wu-book} and references therein.



MAS subjects to actuator saturation has been studied in both semi-global and global frameworks.
Some
researchers have worked on (semi) global state and output
synchronizations for both continuous- and discrete-time MAS subject to actuator
saturation. 
We can summarized the current existing literature on MAS with linear agents subject to actuator saturation as follows:
\begin{enumerate}
	\item The semi-global synchronization has been studied in \cite{su-chen-lam-lin} for
	full-state coupling. For partial state coupling, we have
	\cite{su-chen,zhang-chen-su}
	which are based on the extra communication. 
	Meanwhile, the result without the extra communication is developed in \cite{zhang-saberi-stoorvogel-continues-discrete}. The static protocols via partial state coupling is designed in \cite{liu-saberi-stoorvogel-zhang-ijrnc} for G-passive agents and G-passifiable via input feedforward
	agents. 
	\item Global synchronization for the case of full-state coupling and an undirected network has been studied by
	\cite{yang-meng-dimarogonas-johansson} for neutrally stable and double-integrator discrete-time agents. 
	For special case, the result dealing with networks that has a directed spanning tree are provided in \cite{li-xiang-wei} for single integrator agents. 	
	Then, global synchronization results for partial-state coupling has been
	studied in \cite{liu-saberi-stoorvogel-zhang-ijrnc} for both continuous- and discrete-time G-passive and G-passifiable via input feedforward agent models by a static design, which require the graph is strongly connected and detailed balanced. 		
\end{enumerate}

Recently, we have initiated a research effort on developing scale-free protocols design for MAS, in which the agent model can be continuous- or discrete-time homogeneous and heterogeneous, and includes external disturbances, input delays, and communication delays (\cite{liu-donya-saberi-stoorvogel-TNSE}).  The \emph{scale-free} framework for protocol design
means that the design does not depend on communication network and the size of network or the
number of agents. 
In particular, for the case of linear agents subject to actuator saturation, we have designed the scale-free nonlinear and linear protocols to achieve the global regulated state synchronization for both continuous- or discrete-time agents, see \cite{liu-saberi-stoorvogel-donya-inputsaturation-automatica,liu-donya-saberi-stoorvogel-CCC-7,liu-saberi-stoorvogel-IJRNC-2022}. The nonlinear design is focusing on at most weakly unstable agents, i.e., eigenvalues of agents are in the closed left half plane for continuous-time systems and in the closed unit disc for discrete-time systems. The linear design focused only on the neutrally stable agents.

We continue the research effort on extending the scale-free linear protocol design to global regulated synchronization of discrete-time double-integrator MAS in this paper.
We would like to point out that the double-integrator is polynomially unstable system. It is also well known that the only significant class of MAS with polynomially unstable agents subject to actuator saturation for which global synchronization can be achievable is double-integrator MAS subject to actuator saturation. This class is important due to the practical significance. We propose a family of the linear protocols that are scalable, and any number of this family of protocols can achieve global regulated state synchronization for MAS
with any communication graph with arbitrary number of agents as long as the communication graph has a path between the exosystem which generates reference trajectory and each
agent.



\subsection*{Graphs}
A \emph{weighted graph} $\mathcal{G}$ is defined by a triple
$(\mathcal{V}, \mathcal{E}, \mathcal{A})$ where
$\mathcal{V}=\{1,\ldots, N\}$ is a node set, $\mathcal{E}$ is a set of
pairs of nodes indicating connections among nodes, and
$\mathcal{A}=[a_{ij}]\in \mathbb{R}^{N\times N}$ is the \emph{adjacency}
matrix. Moreover, $a_{ij}=0$ if there is no
edge from node $j$ to node $i$. We assume there are no self-loops,
i.e.\ we have $a_{ii}=0$. A \emph{path} from node $i_1$ to $i_k$ is a
sequence of nodes $\{i_1,\ldots, i_k\}$ such that
$(i_j, i_{j+1})\in \mathcal{E}$ for $j=1,\ldots, k-1$. A
\emph{directed tree} with root $r$ is a subgraph of the graph
$\mathcal{G}$ in which there exists a unique path from node $r$ to
each node in this subgraph. A \emph{directed spanning tree} is a
directed tree containing all the nodes of the graph. The
\emph{weighted in-degree} of node $i$ is given by $d_{\text{in}}(i) = \sum_{j=1}^N\, a_{ij}$. See \cite{saberi-stoorvogel-zhang-sannuti-book}.
For a weighted graph $\mathcal{G}$, the matrix
$L=[\ell_{ij}]$ with
\[
\ell_{ij}=
\begin{system}{cl}
	\sum_{k=1}^{N} a_{ik}, & i=j,\\
	-a_{ij}, & i\neq j,
\end{system}
\]
is called the \emph{Laplacian matrix} associated with the graph
$\mathcal{G}$. The Laplacian matrix $L$ has all its eigenvalues in the
closed right half plane and at least one eigenvalue at zero associated
with right eigenvector $\textbf{1}$. See \cite{royle-godsil}.


\section{System description and problem formulation}
Consider a MAS consisting of $N$ identical discrete-time double-integrator with actuator saturation:
\begin{equation}\label{eq1}
	\begin{cases}
		{x}_i(t+1)=Ax_i(t)+B\sigma(u_i(t)),\\
		y_i(t)=Cx_i(t)
	\end{cases}
\end{equation}
where $x_i(t)\in\mathbb{R}^{2n}$, $y_i(t)\in\mathbb{R}^n$ and $u_i(t)\in\mathbb{R}^n$ are the state, output, and input of agent $i$, respectively,~for $i=1,\ldots, N$.~And
\[
A=\begin{pmatrix}
	I&I\\0&I
\end{pmatrix}, B=\begin{pmatrix}0\\I\end{pmatrix},C=\begin{pmatrix}
	I&0
\end{pmatrix}.
\]
Meanwhile,
\[\sigma(v)=\begin{pmatrix} 
	\sat(v_1) \\\vdots \\ \sat(v_m)
\end{pmatrix},\quad
\text{ where }\quad 
v=\begin{pmatrix} 
	v_1 \\ \vdots \\v_m
\end{pmatrix} \in \R^m
\]
and $\sat(w)$ is standard saturation function satisfying
$
\sat(w)=\sgn(w)\min(1,|w|)
$.

The communication network provides agent $i$ with the following information,
\begin{equation}\label{eq2}
	\zeta_i(t)=\sum_{j=1,i\ne j}^{N}a_{ij}(y_i(t)-y_j(t)),
\end{equation}
where $a_{ij}\geq 0$ and $a_{ii}=0$. The network topology can be described by a weighted graph $\mathcal{G}$ associated with \eqref{eq2}, where
the $a_{ij}$ are the coefficients of adjacency matrix
$\mathcal{A}$. $\zeta_i(t)$ can be rewritten using associated Laplacian matrix as 
\[
\zeta_i(t) = \sum_{j=1}^{N}\ell_{ij}y_j(t).
\]


Now, we consider \textbf{regulated state synchronization} problem. We need a reference trajectory, which is generated by the following exosystem
\begin{align}	
	{x}_r(t+1)&= A x_r(t),\label{solu-cond}\\
	y_r(t)&=Cx_r(t).\label{solu-cond2}	
\end{align}

The objective of regulated state
synchronization is
\begin{equation}\label{synch_org}
\lim_{t\to \infty} (x_{i}(t)-x_r(t))=0
\end{equation}
for all $i \in {1,...,N}$.

Clearly, we need some level of communication between the reference trajectory and the agents. Thus, we assume that a nonempty subset $\mathcal{S}$. The agents in this set have access to their own output relative to the reference trajectory $y_r(t)\in\mathbb{R}^p$. It means that each agent has access to the quantity
\begin{equation}
\psi_i(t)=\iota_i(y_i(t)- y_r(t)), \qquad \iota_i=\begin{system}{cl}
	1, \quad i\in \mathcal{S},\\
	0, \quad i\notin \mathcal{S}.
\end{system}
\end{equation}
Then, the information available for each discrete-time agent can be expressed by
\begin{equation}\label{zeta-bar1}
\bar{\zeta}_i(t)=\frac{1}{2+\bar{D}_{\text{in}}(i)}\sum_{j=1}^N a_{ij}(y_{i}(t)-y_{j}(t))+\iota_i(y_i(t)- y_r(t)),
\end{equation}
where $\bar{D}_{\text{in}}(i)$ is the \emph{upper bound} of $d_{\text{in}}(i)=\sum_{j=1}^N a_{ij}$ for $i=1,\cdots,N$. 
\begin{remark}
As the explanations in \cite[Remark 2]{liu-saberi-stoorvogel-IJRNC-2022}, $\bar{D}_{\text{in}}(i)$ is still a local information. 
\end{remark}

Thus, we use the node set $\mathcal{S}$ as root set. Then, we define an expanded Laplacian matrix as 
\begin{equation}\label{barL}
\bar{L}=L+\diag\{\iota_i\}=[\bar{\ell}_{ij}]_{N\times N}
\end{equation}
for any graph with the Laplacian matrix $L$. Since the sum of its rows does not need to be zero, $\bar{L}$ is not a regular Laplacian matrix associated to the graph. Furthermore, it should be emphasized that $\bar{\ell}_{ij}=\ell_{ij}$ for $i\neq j$ in $\bar{L}$. And \eqref{zeta-bar1} can be rewritten as
\begin{equation}\label{zeta-bar2}
\bar{\zeta}_i(t)=\frac{1}{2+\bar{D}_{\text{in}}(i)}\sum_{j=1}^N\bar{\ell}_{ij}(y_j(t)-y_r(t)).
\end{equation}
To guarantee that each agent can achieve the required regulation, we need to make sure that there exists a path to each node starting with node from the set $\mathcal{S}$. Therefore, we denote the following set of graphs.
\begin{definition}\label{graph-def}
Given a node set $\mathcal{S}$, we denote by $\mathbb{G}_\mathcal{S}^N$ the set of all directed graphs with $N$ nodes containing the node set $\mathcal{S}$, such that every node of the network graph $\mathscr{G}\in \mathbb{G}_\mathcal{S}^N$ is a member of a directed tree which has its root contained in the node set $\mathcal{S}$.
\end{definition}

\begin{remark}
According to Definition \ref{graph-def}, we know that graph set $\mathbb{G}_{\mathcal{S}}^N$ only requires the directed trees and their root in the node set $\mathcal{S}$. Namely, $\mathcal{G}\in\mathbb{G}_{\mathcal{S}}^N$ does not require necessarily the existence of directed spanning tree. 
\end{remark}

For any graph $\mathcal{G}\in \mathbb{G}_\mathcal{S}^N$, with the associated expanded Laplacian matrix $\bar{L}$, we define 
\begin{equation}\label{hodt-LDa}
\bar{D}=I_N-(2I_N+D_{\text{in}})^{-1}{\bar{L}},
\end{equation}
where 
\begin{equation}\label{d-in}
D_{\text{in}}=\diag\{D_{\text{in}}(1),D_{\text{in}}(2),\cdots,D_{\text{in}}(N) \}.
\end{equation}
It is easily verified that the matrix $\bar{D}$ has all eigenvalues in the open unit disk based on \cite[Lemma 1]{liu2018regulated}.


Meanwhile, we introduce an additional
information exchange among each agent and its neighbors. In particular, each agent $i$ ($i=1,\ldots, N$) has access to the following additional information
$\hat{\zeta}_i(t)$, i.e.
\begin{equation}\label{eqa1}
\hat{\zeta}_i(t)=\frac{1}{2+\bar{D}_{\text{in}}(i)}\sum_{j=1}^Na_{ij}(\xi_i(t)-\xi_j(t))
\end{equation}
where $\xi_j(t)$ is a variable produced internally by agent $j$ and defined in the following sections.

Then, we formulate the global regulated state synchronization problem for a MAS via
linear protocols based on the additional information exchange \eqref{eqa1}.
\begin{problem}\label{prob4}
Consider a MAS described by \eqref{eq1}, and
the associated exosystem \eqref{solu-cond} and \eqref{solu-cond2}. Let a set of nodes
$\mathcal{S}$ be given which defines the set
$\mathbb{G}_{\mathcal{S}}^N$. Let the associated network
communication be given by Definition \ref{graph-def}.

The \textbf{scalable global regulated state synchronization problem
	with additional information exchange via linear	dynamic protocol} is to find a linear dynamic protocol,
using only the knowledge of agent model $(A,B,C)$, of the form 
\begin{equation*}
	\begin{system}{cll}
		x_{c,i}(t+1)&=&A_{c,i} x_{c,i}(t)+B_{c,i}{\sigma(u_i(t))}\\
		&&\hspace{1cm}+C_{c,i}
		\bar{\zeta}_i(t)+D_{c,i} \hat{\zeta}_i(t),\\
		u_i(t)&=&K_{c,i}x_{c,i}(t),
	\end{system}
\end{equation*}
where $\hat{\zeta}_i(t)$ is defined in \eqref{eqa1} with
$\xi_i(t)=H_{c,i}x_{c,i}(t)$, and $x_{c,i}(t)\in\R^{n_c}$, such that regulated
state synchronization \eqref{synch_org} is achieved for any $N$ and any
graph $\mathcal{G}\in \mathbb{G}_{\mathcal{S}}^N$, and for all
initial conditions of the agents $x_i(0) \in \mathbb{R}^n$, the exosystem $x_r(0) \in \mathbb{R}^n$, and the protocols
$x_{c,i}(0) \in \mathbb{R}^{n_c}$.
\end{problem}


\section{Scale-free protocol design}
\subsection{Full-state coupling case (i.e., $C=I$)}
We design the following linear dynamic protocol. 
\begin{tcolorbox}[breakable,colback=white]
\begin{equation}\label{pscpd1}
	\begin{system}{rl}
		{\chi}_i(t+1) =&
		A\chi_i(t)+B\sigma(u_i(t))+ 
		A\bar{\zeta}_i(t)\\
		&\hspace{0.5cm}-A\hat{\zeta}_i(t)
		-\frac{\iota_i}{2+\bar{D}_{\text{in}}(i)} A\chi_i(t), \\
		u_i(t) =&-K\chi_i(t).
	\end{system}
\end{equation}
Then, we choose matrix $K=\begin{pmatrix}
	k_1 I&k_2 I
\end{pmatrix}$, where $k_1$ and $k_2$ satisfy the following condition
\begin{equation}\label{cond1}
	\begin{system}{l}
		0<k_1<2,\\
		k_2>0,\\
		(4+k_1-2k_2)(3k_1-2k_2)<0.
	\end{system}	
\end{equation}
And the agents communicate $\xi_i(t)$ which is chosen as $\xi_i(t)=\chi_i(t)$. Namely,
\begin{equation}\label{add_1f}
	\hat{\zeta}_{i}(t)=\frac{1}{2+\bar{D}_{\text{in}}(i)}\sum_{j=1}^N a_{ij}(\chi_i(t)-\chi_j(t)).
\end{equation}
$\bar{\zeta}_i(t)$ is defined
by \eqref{zeta-bar1}.
\end{tcolorbox}
\begin{remark}
	Please note that since agent $i$ has access to $\psi_i(t)$, that implies $\iota_i$ is known to agent $i$.
\end{remark}

\begin{figure}[h!]
\includegraphics[width=7cm]{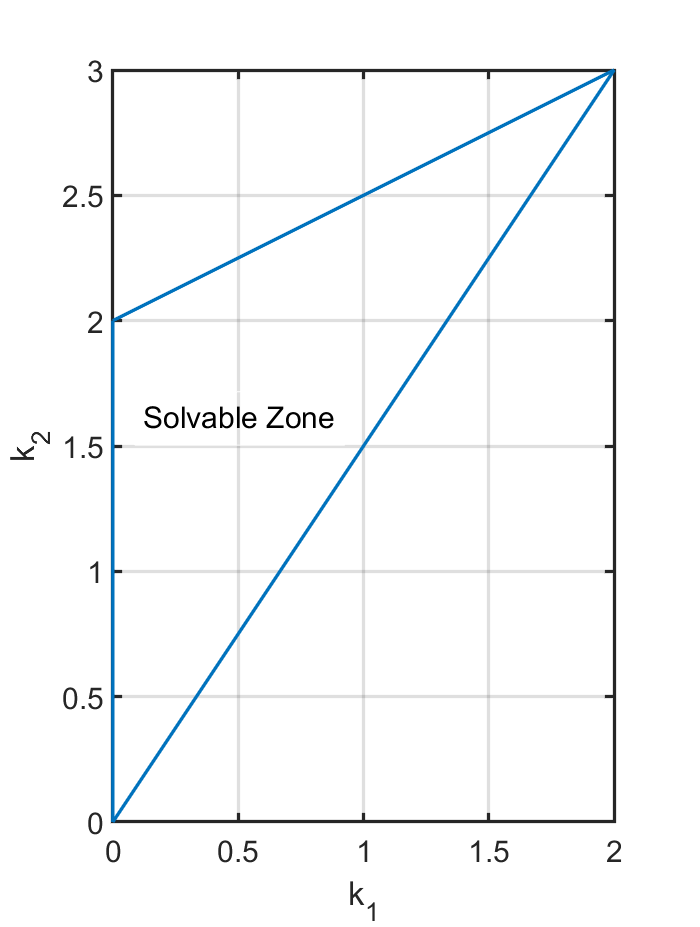}
\centering
\caption{Solvable zone of $k_1, k_2$ for global synchronization}\label{zone}
\end{figure}

The condition \eqref{cond1} can be shown as Fig. \ref{zone}, where the triangular formed by three points $(0,0)$, $(0,2)$, and $(2,3)$ is the solvable zone of $k_1,k_2$.
\begin{remark}
It is interesting to point out that for discrete-time double-integrator system subject to saturation, we obtain global stablization by linear state feedback law and the feedback gain belongs to the exactly same set as given the zone for parameter $k_1, k_2$, see \cite{yang-stoorvogel-saberi2}, \cite{yang-stoorvogel-saberi-discrete-journal}, \cite{yang-stoorvogel-wang-saberi}.
\end{remark}

Then, we have the following theorem by using the above design.
\begin{theorem}\label{mainthm3}
Consider a MAS described by \eqref{eq1} with $C=I$, and the associated exosystem
\eqref{solu-cond}. Let a set of nodes $\mathcal{S}$ be given which
defines the set $\mathbb{G}_{\mathcal{S}}^N$. Let the associated
network communication be given by \eqref{zeta-bar1}. 

Then, the \textbf{scalable global regulated state synchronization problem with additional information exchange} as stated in Problem
\ref{prob4} is solvable. In particular, for any given $k_1$ and $k_2$ satisfying \eqref{cond1}, the linear
dynamic protocol \eqref{pscpd1} solves the global regulated state
synchronization problem for any $N$ and any graph
$\mathcal{G}\in\mathbb{G}_{\mathcal{S}}^N$. 
\end{theorem}

To obtain this theorem we need the following lemma.
\begin{lemma}\label{lem1} For all $u,v\in \R^n$, we have
\begin{equation}\label{cond0}
	(\sigma(v)-\sigma(u))\T(u-\sigma(u))\leq 0.
\end{equation}
\end{lemma}
\begin{proof}
Note that we have:
\begin{equation}\label{lemc1}
	(\sigma(v)-\sigma(u))\T(u-\sigma(u))  =\sum_{i=1}^n
	(\sigma(v_i)-\sigma(u_i))(u_i-\sigma(u_i))
\end{equation}
when $u=\begin{pmatrix} u_1 & \cdots & u_n \end{pmatrix}\T$ and $v=\begin{pmatrix} v_1 & \cdots & v_n \end{pmatrix}\T$.
Next note that if $u_i\geq 1$ we have
$\sigma(v_i)-\sigma(u_i)=\sigma(v_i)-1\leq0$ and
$u_i-\sigma(u_i)=u_i-1\geq 0$ and hence:
\begin{equation}\label{lemc2}
	(\sigma(v_i)-\sigma(u_i))(u_i-\sigma(u_i))\leq 0
\end{equation}
On the other hand if $u_i\leq -1$ we have
$\sigma(v_i)-\sigma(u_i)=\sigma(v_i)+1\geq0$ and
$u_i-\sigma(u_i)=u_i+1\leq 0$ and \eqref{lemc2} is still
satisfied.
Finally, if  $|u_i|\leq 1$ then $u_i-\sigma(u_i)=0$ and
\eqref{lemc2} is also satisfied.

Since \eqref{lemc2} is satisfied for all $i$ and using
\eqref{lemc1} we find \eqref{cond0} holds for all $u$ and $v$.
\end{proof}

\begin{proof}[The proof of Theorem \ref{mainthm3}]
Firstly, let $\tilde{x}_i(t)=x_i(t)-x_r(t)$, we have
\begin{equation}
	\begin{system}{l}
		\tilde{x}_{i}(t+1)=A\tilde{x}_{i}(t)+B\sigma(u_i(t)),\\
		\chi_i(t+1)=A\chi_i(t)+B\sigma(u_i(t))\\
		\hspace{2cm}+ \frac{1}{2+\bar{D}_{\text{in}}(i)}
		\sum_{j=1}^N\bar{\ell}_{ij} A\left[\tilde{x}_{j}(t)-\chi_j(t) \right], \\
		u_i(t)=-\begin{pmatrix}k_1 I\quad&k_2  I\end{pmatrix}\chi_i(t),
	\end{system}
\end{equation}
where 
$
\frac{1}{2+\bar{D}_{\text{in}}(i)}\sum_{j=1}^N\bar{\ell}_{ij} A\chi_j(t)
=A\hat{\zeta}_i(t)+\frac{\iota_i}{2+\bar{D}_{\text{in}}(i)} A\chi_i(t)$.
Then by defining 
\begin{multline*}
	\tilde{x}(t)=\begin{pmatrix}
		\tilde{x}_1(t)\\\vdots\\\tilde{x}_N(t)
	\end{pmatrix},   
	\chi(t)=\begin{pmatrix}
		\chi_1(t)\\\vdots\\\chi_N(t)
	\end{pmatrix}, \\ u(t)=\begin{pmatrix}
		u_1(t)\\\vdots\\u_N(t)
	\end{pmatrix},\sigma(u(t))=\begin{pmatrix}
		\sigma(u_1(t))\\\vdots\\\sigma(u_N(t))
	\end{pmatrix},
\end{multline*}
one can obtain the closed-loop system in the form of
\begin{equation}
	\begin{system}{l}
		\tilde{x}(t+1)=(I_N\otimes A)\tilde{x}(t)+(I_N\otimes B)\sigma(u(t))\\
		\chi(t+1)=(I_N\otimes A)\chi(t)+(I_N\otimes B)\sigma(u(t))\\
		\hspace{2cm}+((I_N-\bar{D})\otimes A)(\tilde{x}(t)-\chi(t))\\
		u(t)=-\left(I_N\otimes \begin{pmatrix}k_1 I\quad&k_2 I\end{pmatrix}\right)\chi(t).
	\end{system}
\end{equation}
Let $e(t)=\tilde{x}(t)-\chi(t)$, we have
\begin{equation}\label{eqn1}
	\begin{system}{l}
		\tilde{x}(t+1)=(I_N\otimes A)\tilde{x}(t)+(I_N\otimes B)\sigma(u(t)),\\
		e(t+1)=(\bar{D}\otimes A)e(t),\\
		u(t)=-\left(I_N\otimes \begin{pmatrix}k_1 I\quad&k_2 I\end{pmatrix}\right)(\tilde{x}(t)-e(t)).
	\end{system}
\end{equation}
Then, let
\[
\tilde{x}_{I}(t)=
\left(I_N\otimes \begin{pmatrix}I& 0\end{pmatrix}\right)\tilde{x}(t)
\text{ and }
\tilde{x}_{II}(t)
=\left(I_N\otimes  \begin{pmatrix}0& I\end{pmatrix}
\right)\tilde{x}(t),
\]
we have
\begin{equation}\label{eqo1}
	\begin{system}{l}
		\tilde{x}_{I}(t+1)=\tilde{x}_{I}(t)+\tilde{x}_{II}(t)\\
		\tilde{x}_{II}(t+1)=\tilde{x}_{II}(t)+\sigma(u(t))\\
		e(t+1)=(\bar{D}\otimes A)e(t)\\
		u(t)=-k_1 \tilde{x}_{I}(t)-k_2 \tilde{x}_{II}(t)+\left(I_{N}\otimes \begin{pmatrix}k_1 I\quad&k_2 I\end{pmatrix}\right)e(t).
	\end{system}
\end{equation}

The eigenvalues of $\bar{D}\otimes A$ are of the form
$\lambda_i \eta_j$, with $\lambda_i$ and $\eta_j$ eigenvalues of
$\bar{D}$ and $A$, respectively. Since $|\lambda_i|<1$ and
$\eta_j\equiv 1$, we find $\bar{D}\otimes A$ is Schur
stable. Therefore we find that 
\begin{equation}\label{estable}
	\lim_{t\to \infty}e_i(t)\to 0.
\end{equation}
It also shows that $e_i\in \ell_2$.  Thus, we just need to prove
the stability of \eqref{eqo1}. Namely, we have $\tilde{x}(t)\to 0$ as $t\to \infty$ with $e_i\in \ell_2$, which means that the synchronization result is obtain.

Then, we consider the following weighting Lyapunov function
\begin{equation}\label{lyapunvd1}
	V(t)=(1-h)V_1(t)+hV_2(t),
\end{equation}
where $h\in(0,1)$,
\begin{align*}
	V_1(t)=&\begin{pmatrix}
		\sigma(u(t))\\\tilde{x}_{II}(t)
	\end{pmatrix}\T
	\left[\begin{pmatrix}
		1+\frac{k_1}{2}& k_1\\ k_1&{k_1}
	\end{pmatrix}\otimes I_{Nn}\right]
	\begin{pmatrix}
		\sigma(u(t))\\\tilde{x}_{II}(t)
	\end{pmatrix}\\
	&\hspace{2cm}+2\sigma(u(t))\T( u(t)-\sigma(u(t))),\\
	V_2(t)=&e\T(t) P_De(t)
\end{align*}
with $P_D>0$ satisfying 
\begin{equation}\label{condd3}
	(\bar{D}\otimes A)\T P_D(\bar{D}\otimes A)- P_D\leq -2I_{2Nn}.
\end{equation}
Here, we obtain $V_1(t)$ and $V_2(t)$ are positive due to $0<k_1<2$ and $P_D>0$, i.e., $V_1(t)>0$ except for $(u(t), \tilde{x}_{II}(t))=0$ when $V_1(t)=0$ and $V_2(t)>0$ except for $e(t)=0$ when $V_2(t)=0$. 
Then, we have
\begin{align*}
	&\Delta V_1(t)=V_1(t+1)-V_1(t)\\
	=&-(1-\frac{k_1}{2})\sigma(u(t+1))\T \sigma(u(t+1))+2\sigma(u(t+1))\T u(t)\\
	&+2(k_1-k_2) \sigma(u(t+1))\T\sigma(u(t))\\		
	&+(1+\frac{k_1}{2}) \sigma(u(t))\T\sigma(u(t))-2\sigma(u(t))\T u(t)\\
	&+2\sigma(u(t+1))\T(I_{N-1}\otimes (k_1 I \quad k_2 I)\Psi)e(t)\\
	=&2(\sigma(u(t+1))-\sigma(u(t)))\T(u(t)-\sigma(u(t)))\\
	&+2(1+k_1-k_2) \sigma(u(t+1))\T\sigma(u(t))\\
	&-(1-\frac{k_1}{2})\left(\sigma(u(t+1))\T \sigma(u(t+1))+\sigma(u(t))\T\sigma(u(t))\right)\\
	&+2\sigma(u(t+1))\T(I_{N-1}\otimes (k_1 I \quad k_2 I)\Psi)e(t)\\
	\leq&2(1+k_1-k_2) \sigma(u(t+1))\T\sigma(u(t))\\
	&-(1-\frac{k_1}{2})\left(\sigma(u(t+1))\T \sigma(u(t+1))+\sigma(u(t))\T\sigma(u(t))\right)\\
	&+2\sigma(u(t+1))\T(I_{N-1}\otimes (k_1 I \quad k_2 I)\Psi)e(t)%
\end{align*}
since $(\sigma(u(t+1))-\sigma(u(t)))\T(u(t)-\sigma(u(t)))\leq 0$ based on Lemma \ref{lem1}, where $\Psi=\bar{D}\otimes A-I_{2Nn}$.
Meanwhile, for $V_2(t)$, we have
\begin{equation*}
	\Delta V_2(t)=V_2(t+1)-V_2(t)\leq -2e\T(t) e(t)
\end{equation*}
based on condition \eqref{condd3}. 	
Thus, one can obtain
\begin{align*}
	&\Delta V(t)\leq (1-h)\Delta V_1(t)+h\Delta V_2(t)\\
	\leq &2(1-h)(1+k_1-k_2) \sigma(u(t+1))\T\sigma(u(t))\\
	&-(1-h)(1-\frac{k_1}{2}-\frac{\|\Psi\|^2(1-h)(k_1^2+k_2^2)}{h})\|\sigma(u(t+1))\|^2 \\
	&-(1-h)(1-\frac{k_1}{2})\sigma(u(t))\T\sigma(u(t))\\
	\leq&(1-h)\begin{pmatrix}
		\sigma(u(t+1))\\\sigma(u(t))
	\end{pmatrix}\T (\Phi\otimes I_{Nn} ) \begin{pmatrix}
		\sigma(u(t+1))
		\\\sigma(u(t))
	\end{pmatrix}-h\|e(t)\|^2,
\end{align*}	
where 
\begin{equation}\label{cond2}
	\Phi=\begin{pmatrix}
		-1+\frac{k_1}{2}+\frac{\|\Psi\|^2(1-h)(k_1^2+k_2^2)}{h}&1+k_1-k_2\\
		1+k_1-k_2&-1+\frac{k_1}{2}
	\end{pmatrix}.
\end{equation}	
Obviously we just need to prove $\Phi< 0$. 

Meanwhile, we have 
\[
(4+k_1-2k_2)(3k_1-2k_2)=4(1+k_1-k_2)^2-4(1-\frac{k_1}{2})^2<0,
\]
and $0<k_1<2$. 
Without loss of generality, thus there exists an $\eps>0$ such that 
\begin{equation}\label{cond3}
	\frac{(1+k_1-k_2)^2}{1-\frac{k_1}{2}}= 1-\frac{k_1}{2}-\eps.
\end{equation}

Using Schur Compliment, we have $\Phi< 0$ is equivalent to
\[
-1+\frac{k_1}{2}+\frac{\|\Psi\|^2(1-h)(k_1^2+k_2^2)}{h}+\frac{(1+k_1-k_2)^2}{1-\frac{k_1}{2}}< 0.
\]
According to condition \eqref{cond3}, we can obtain
\begin{align*}
	&-1+\frac{k_1}{2}+\frac{\|\Psi\|^2(1-h)(k_1^2+k_2^2)}{h}+\frac{(1+k_1-k_2)^2}{1-\frac{k_1}{2}}\\
	< &\frac{\|\Psi\|^2(1-h)(k_1^2+k_2^2)}{h}-\eps.
\end{align*}
Let $h$ sufficiently close to 1, the above inequality is less than zero, i.e., 
\[
\frac{\|\Psi\|^2(1-h)(k_1^2+k_2^2)}{h}<\eps.
\]
It also means that $\Phi< 0$.	
Thus, we have $\Delta V(t)< 0$ for $\begin{pmatrix}
	\sigma(u(t+1))
	\\\sigma(u(t))
\end{pmatrix}\neq 0$, $\tilde{x}(t)\to 0$ as $t\to \infty$. Furthermore, when $\Delta V(t)=0$, we obtain $u(t+1)=u(t)=0$ and $e(t)=0$. It is easy to obtain $\tilde{x}_{I}(t)=\tilde{x}_{II}(t)=0$ at $\Delta V(t)=0$. 

Thus, \eqref{eqn1} is globally asymptotically stable based on LaSalle’s invariance principle, i.e. the scale-free global regulated state synchronization result can be obtained via protocol \eqref{pscpd1} with condition \eqref{cond1}.	
\end{proof}

\subsection{Partial-state coupling case (i.e., $C\ne I$)}
We design a linear dynamic protocol with additional information
exchanges \eqref{eqa1} via partial-state coupling for agent $i$ ($i\in\{1,\ldots,N\}$).
\begin{tcolorbox}[breakable,colback=white]
\begin{equation}\label{pscpd2}
	\begin{system}{rl}
		\hat{x}_i(t+1) &=
		(A-FC)\hat{x}_i(t) + B\hat{\zeta}_{i2}(t)\\
		&\hspace{2cm}+F\bar{\zeta}_i(t)+ \frac{\iota_i}{2+\bar{D}_{\text{in}}(i)} 
		B\sigma(u_i(t)) \\ 
		{\chi}_i(t+1) &= A\chi_i(t)+B\sigma(u_i(t))+A\hat{x}_i(t)\\
		&\hspace{2cm} -	 A\hat{\zeta}_{i1}(t) - \frac{\iota_i}{2+\bar{D}_{\text{in}}(i)} A\chi_i(t) \\
		u_i(t) &=  -K \chi_i(t).
	\end{system}
\end{equation}
Then, we choose matrix $K=\begin{pmatrix}
	k_1 I&k_2 I
\end{pmatrix}$, where $k_1,k_2$ satisfy condition \eqref{cond1}. 		
In this protocol, the agents communicate 
	\[
	\xi_i(t)=\begin{pmatrix} \xi_{i1}(t) \\ \xi_{i2}(t) 
	\end{pmatrix}=\begin{pmatrix}
		\chi_i(t) \\ \sigma (u_i(t))
	\end{pmatrix},
	\]
	i.e.,\ each agent has access to additional information 
		\[
		\hat{\zeta}_i(t)=\begin{pmatrix} 
			\hat{\zeta}_{i1}(t) \\ \hat{\zeta}_{i2} (t)
		\end{pmatrix},
		\]
		where
		\begin{equation}\label{add_1}
			\begin{system}{l}
				\hat{\zeta}_{i1}(t)=\frac{1}{2+\bar{D}_{\text{in}}(i)}\sum_{j=1}^N a_{ij}(\chi_i(t)-\chi_j(t)) \\
				\hat{\zeta}_{i2}(t)=\frac{1}{2+\bar{D}_{\text{in}}(i)}\sum_{j=1}^{N} a_{ij}(\sigma (u_i(t))-\sigma (u_j(t)))
			\end{system}		
		\end{equation}
		and $\bar{\zeta}_i(t)$ is defined in \eqref{zeta-bar1}.
	\end{tcolorbox}
	
	We propose the following theorem.
	\begin{theorem}\label{mainthm4}
		Consider a MAS described by \eqref{eq1}, and the associated exosystem
		\eqref{solu-cond} and \eqref{solu-cond2}. Let a set of nodes $\mathcal{S}$ be given which
		defines the set $\mathbb{G}_{\mathcal{S}}^N$. Let the associated
		network communication be given by \eqref{zeta-bar1}. 
		
		Then, the \textbf{scalable global regulated state synchronization problem with additional information exchange} as stated in Problem
		\ref{prob4} is solvable. In particular, for any given $k_1$ and $k_2$ satisfying \eqref{cond1}, the linear
		dynamic protocol \eqref{pscpd2} solves the global regulated state
		synchronization problem for any $N$ and any graph
		$\mathcal{G}\in\mathbb{G}_{\mathcal{S}}^N$. 
	\end{theorem}
	
	\begin{proof}[The proof of Theorem \ref{mainthm4}]
		Similar to Theorem \ref{mainthm3}, we have
		the matrix expression of closed-loop system by defining
		$\tilde{x}_i(t)=x_i(t)-x_r (t)$, 
		\begin{multline*}
			\tilde{x}(t)=\begin{pmatrix}
				\tilde{x}_1(t)\\\vdots\\\tilde{x}_N(t)
			\end{pmatrix},   
			\chi(t)=\begin{pmatrix}
				\chi_1(t)\\\vdots\\\chi_N(t)
			\end{pmatrix}, \\ u(t)=\begin{pmatrix}
				u_1(t)\\\vdots\\u_N(t)
			\end{pmatrix},\sigma(u(t))=\begin{pmatrix}
				\sigma(u_1(t))\\\vdots\\\sigma(u_N(t))
			\end{pmatrix},
		\end{multline*}
		$\tilde{x}_{I}(t)=
		\left(I_N\otimes \begin{pmatrix}I& 0\end{pmatrix}\right)\tilde{x}(t)$, 
		$\tilde{x}_{II}(t)
		=\left(I_N\otimes  \begin{pmatrix}0& I\end{pmatrix}
		\right)\tilde{x}(t)$,
		$e(t)=\tilde{x}(t)-\chi(t)$, and
		$\bar{e}(t)=[(I_N-\bar{D})\otimes I]\tilde{x}(t)-\hat{x}(t)$. Then, the closed-loop system is expressed as follows,
		\begin{equation*}\label{newsysted}
			\begin{system}{l}
				\tilde{x}_{I}(t+1)=\tilde{x}_{I}(t)+\tilde{x}_{II}(t)\\
				\tilde{x}_{II}(t+1)=\tilde{x}_{II}(t)+\sigma(u(t))\\
				e(t+1)=(\bar{D}\otimes A)e(t)+\bar{e}(t)\\
				\bar{e}(t+1)=[I_{N}\otimes (A-FC)]\bar{e}(t)\\
				u(t)=-\left(I_{N}\otimes \begin{pmatrix}k_1 I\quad&k_2 I\end{pmatrix}\right)\chi(t).
			\end{system}
		\end{equation*}  
		Since the eigenvalues of $A-FC$ and $\bar{D}\otimes A$ are in open unit disk, we just need to prove 
		the stability of $\tilde{x}_{I}(t)$ and $\tilde{x}_{II}(t)$.

		Similar to the proof of Theorem \ref{mainthm3}, we can obtain the scale-free global regulated state synchronization
		result.
	\end{proof}

	\section{Numerical examples}
	We use three numerical examples to illustrate the effectiveness and scalability of our designs for discrete-time double-integrator MAS. 
	
	Firstly, we give the agent models \eqref{eq1}, exosystem model \eqref{solu-cond}, and \eqref{solu-cond2} with the following parameters 
	\[
	A=\begin{pmatrix}
		1&1\\0&1
	\end{pmatrix}, B=\begin{pmatrix}0\\1\end{pmatrix},C=\begin{pmatrix}
		1&0
	\end{pmatrix}.
	\]
	
	To show the scalability of our protocol design, we use two graphs which consists of 6 and 60 nodes (or agents). We consider the case of partial-state coupling and assume that only Agent 1 can receive the exosystem's signal, i.e. $\iota_1=1$. To show the effectiveness of our protocol design based on condition \eqref{cond1}, we choose 3 groups of parameters ($k_1, k_2$): 
	\begin{enumerate}
		\item $k_1=0.5$ and $k_2=1$.
		\item $k_1=1$ and $k_2=2$.
		\item $k_1=1.5$ and $k_2=2.5$.
	\end{enumerate}
	The detailed protocol is provided as follows:
	\begin{equation}\label{protoclsim}
		\begin{system}{ll}
			\hat{x}_i(t+1) &=
			\begin{pmatrix}
				-0.5&1\\-0.5&1
			\end{pmatrix}\hat{x}_i(t)
			+  \begin{pmatrix}
				0\\1
			\end{pmatrix}\hat{\zeta}_{i2}(t)\\
			&\hspace{2cm}+ \frac{\iota_i}{2+\bar{D}_{\text{in}}(i)} 
			\begin{pmatrix}
				0\\1
			\end{pmatrix}\sigma(u_i(t))+\begin{pmatrix}
				1.5\\0.5
			\end{pmatrix}\bar{\zeta}_i(t) \\ 
			{\chi}_i(t+1) &= \begin{pmatrix}
				1&1\\0&1
			\end{pmatrix}\left[\chi_i(t)-
			( \hat{\zeta}_{i1}(t) + \frac{\iota_i}{2+\bar{D}_{\text{in}}(i)} \chi_i(t) )\right]\\
			&\hspace{2cm}+\begin{pmatrix}
				0\\1
			\end{pmatrix}\sigma(u_i(t))+\begin{pmatrix}
				1&1\\0&1
			\end{pmatrix}\hat{x}_i(t)\\
			u_i(t) &=  -\begin{pmatrix}
				k_1 &k_2 
			\end{pmatrix} \chi_i(t).
		\end{system}
	\end{equation}

	
	Then, we use three communication networks consisting of 3, 6, and 60 agents respectively, and with different communication topologies. 
	These two cases show that protocol design is scale-free and does not depend on the communication network and the number of
	agents $N$. 
	
	
	\begin{figure}[ht!]
		\includegraphics[width=8cm]{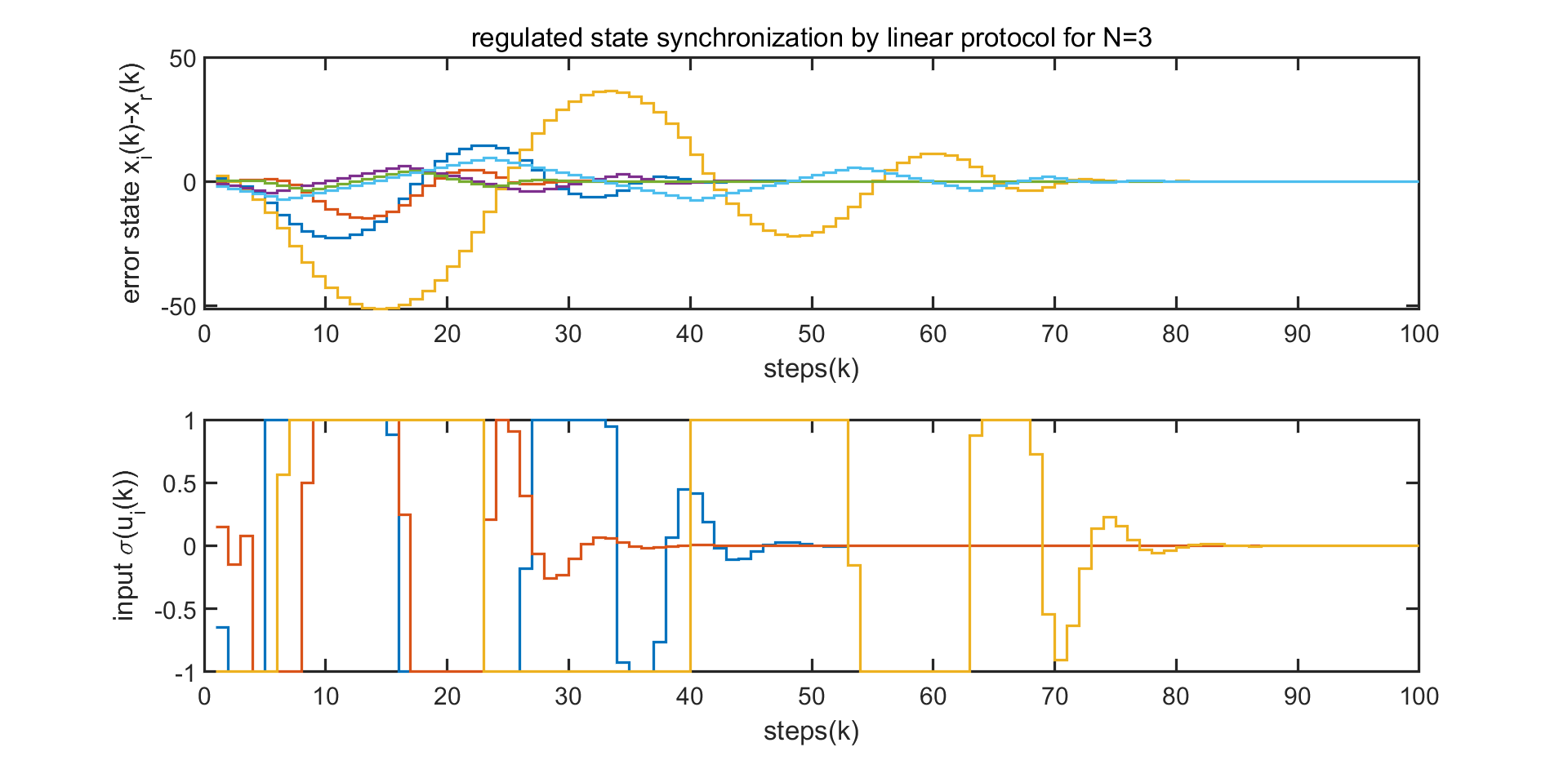} \centering
		\caption{Regulated state synchronization for
			MAS with communication graph
			I when $k_1=0.5$ and $k_2=1$.}\label{results_case1.1}
	\end{figure}
	\begin{figure}[ht!]
		\includegraphics[width=8cm]{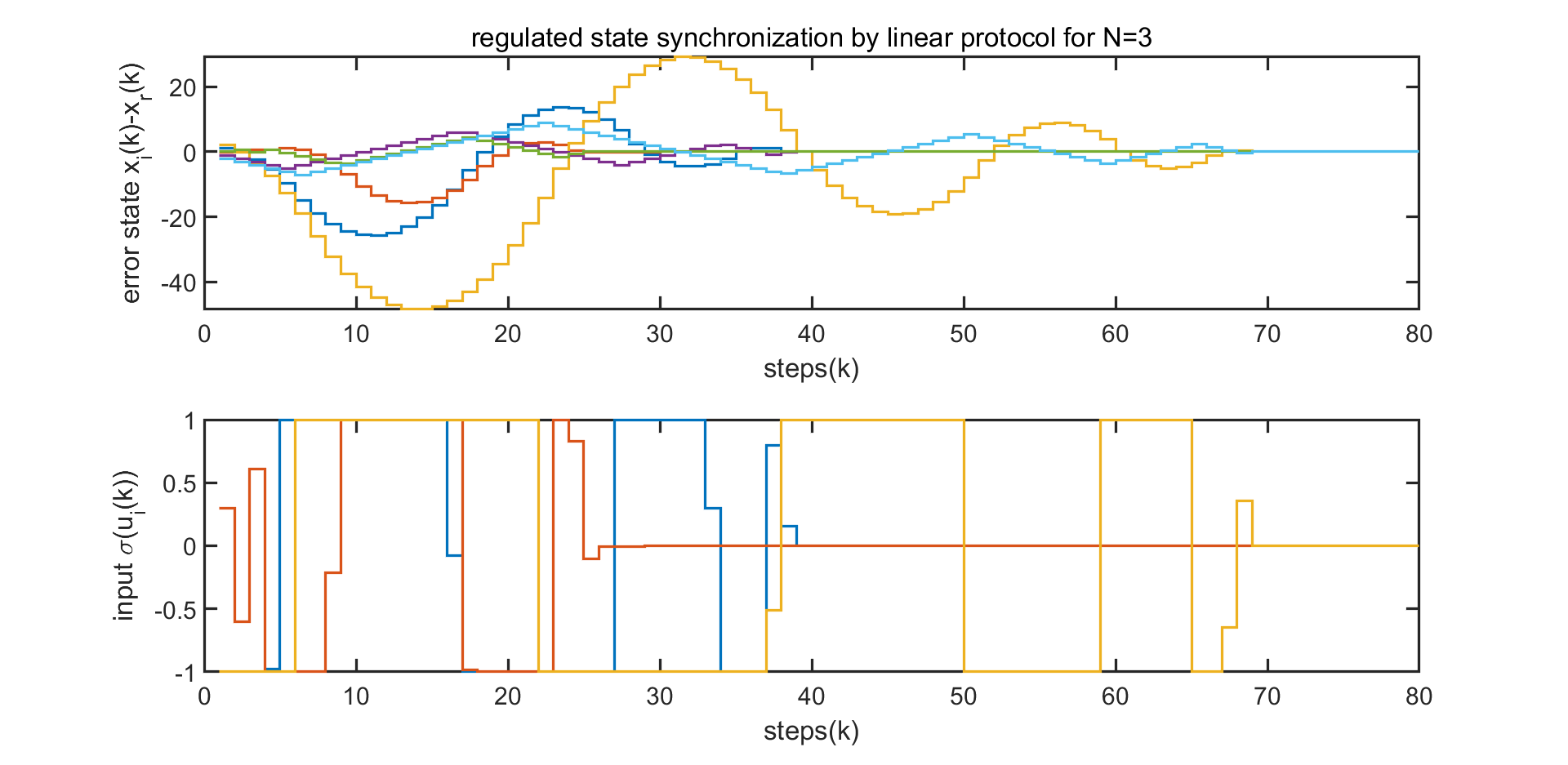} \centering
		\caption{Regulated state synchronization for
			MAS with communication graph
			I when $k_1=1$ and $k_2=2$.}\label{results_case1.2}
	\end{figure}
	\begin{figure}[ht!]
		\includegraphics[width=8cm]{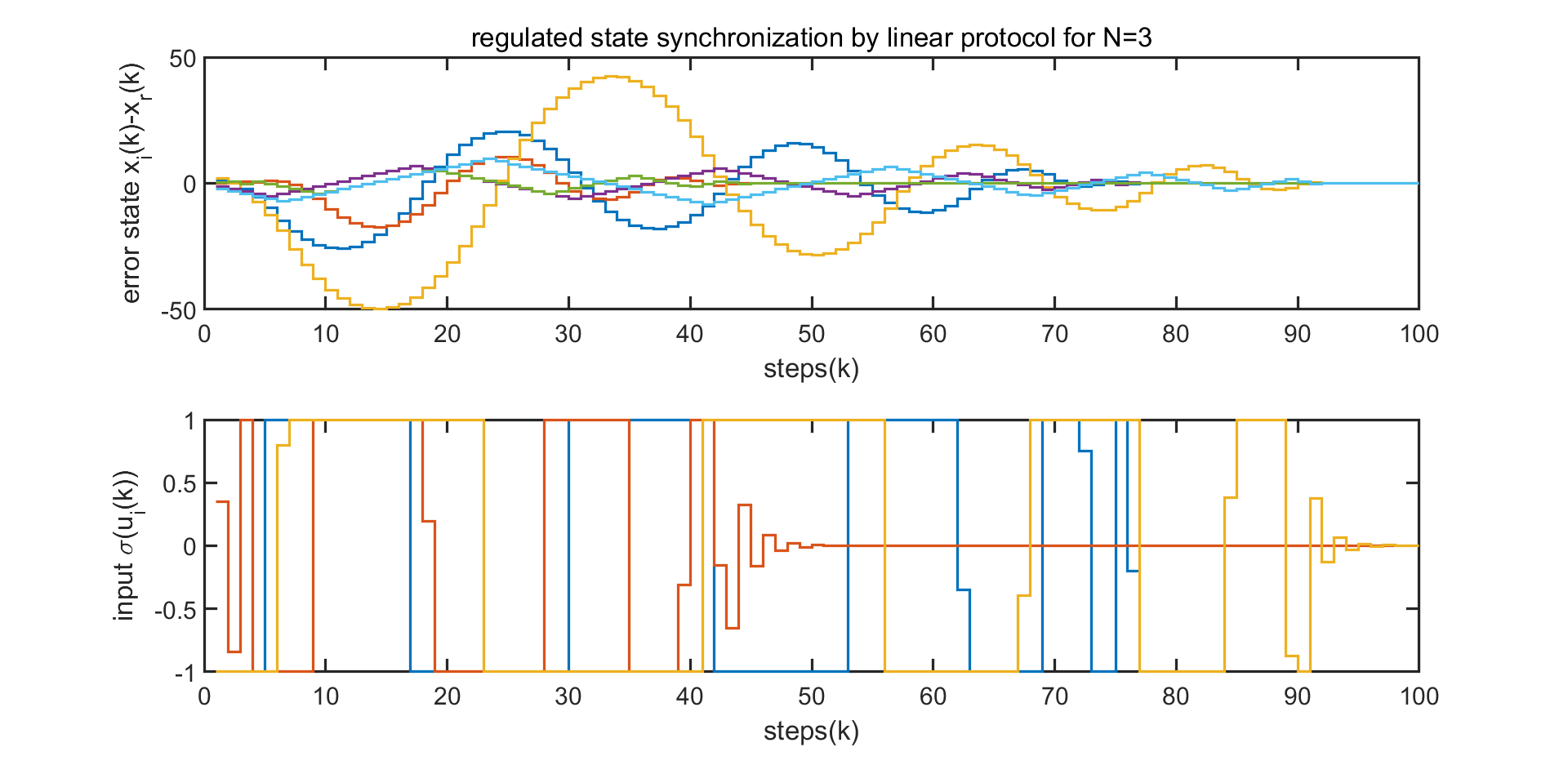} \centering
		\caption{Regulated state synchronization for
			MAS with communication graph
			I when $k_1=1.5$ and $k_2=2.5$.}\label{results_case1.3}
	\end{figure}
	
	\subsection*{\bf Case I: 3-agent graph}
	In this case, we consider a MAS with $3$ agents,
	$N=3$. The associated adjacency matrix to the communication network is assumed to be $\mathcal{A}_I$ where $a_{21}=a_{32}=1$. 
	
	The simulation results of Case I by using protocol \eqref{protoclsim} are demonstrated in Figure
	\ref{results_case1.1}--\ref{results_case1.3}. The results show that the
	protocol design is effective for different ($k_1, k_2$) satisfying condition \eqref{cond1}.

	\begin{figure}[ht!]
		\includegraphics[width=8cm]{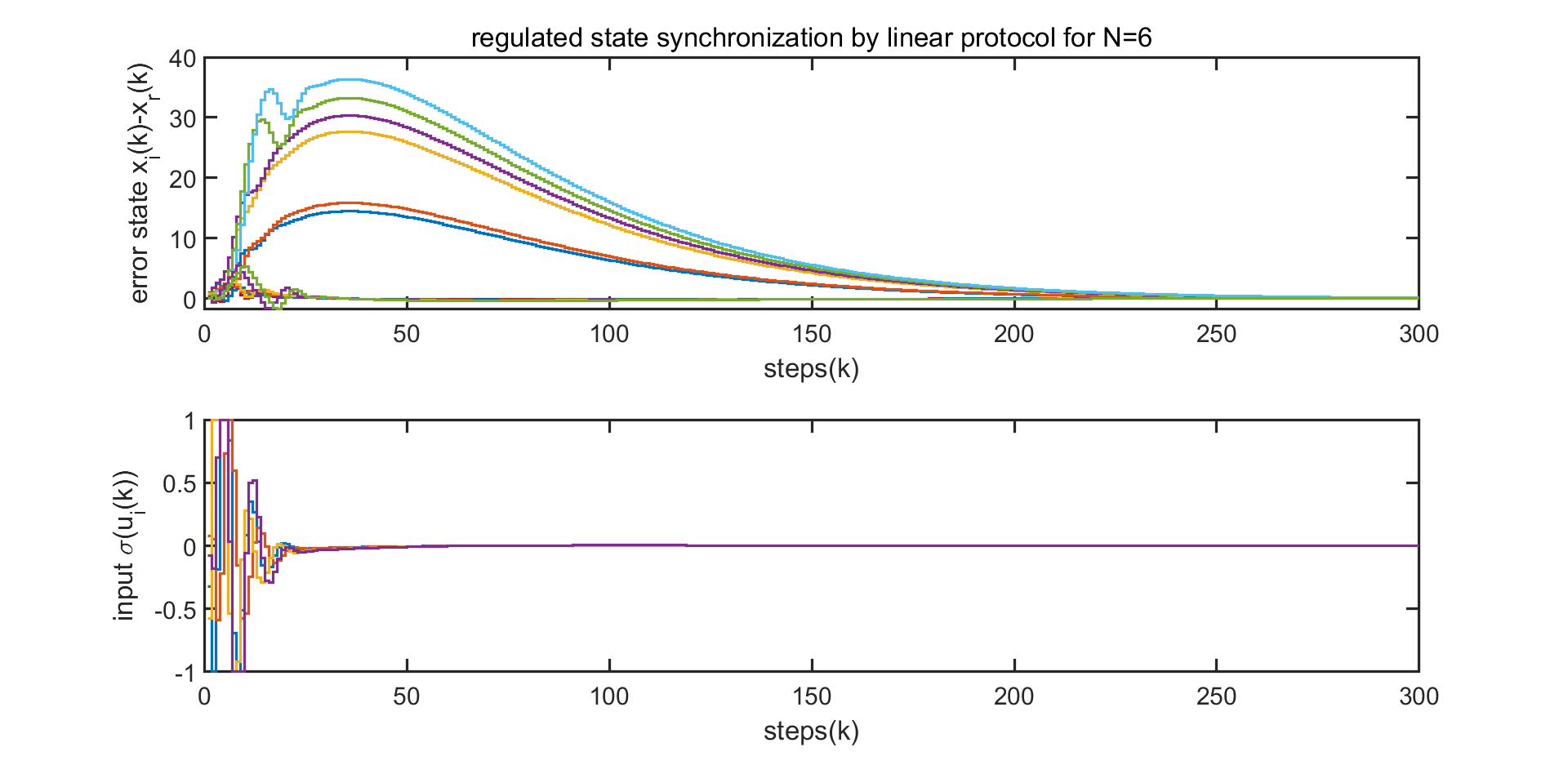} 
		\centering
		\caption{{Regulated state synchronization for
				MAS with communication graph
				II when $k_1=0.5,k_2=1$.}}\label{results_case2.1}
	\end{figure}
	\begin{figure}[ht!]
		\includegraphics[width=8cm]{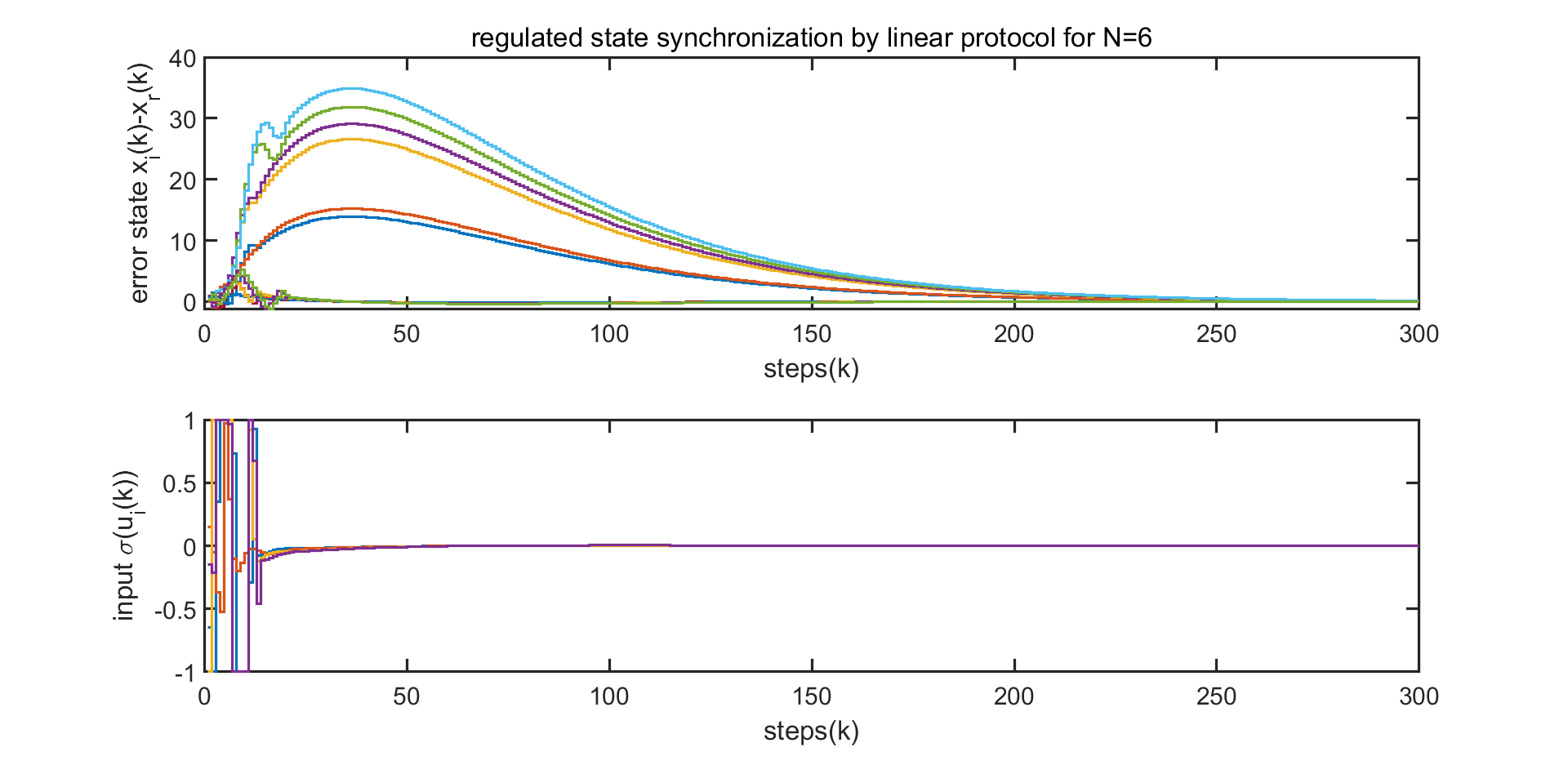} \centering
		\caption{Regulated state synchronization for
			MAS with communication graph
			II when $k_1=1$ and $k_2=2$.}\label{results_case2.2}
	\end{figure}
	\begin{figure}[ht!]
		\includegraphics[width=8cm]{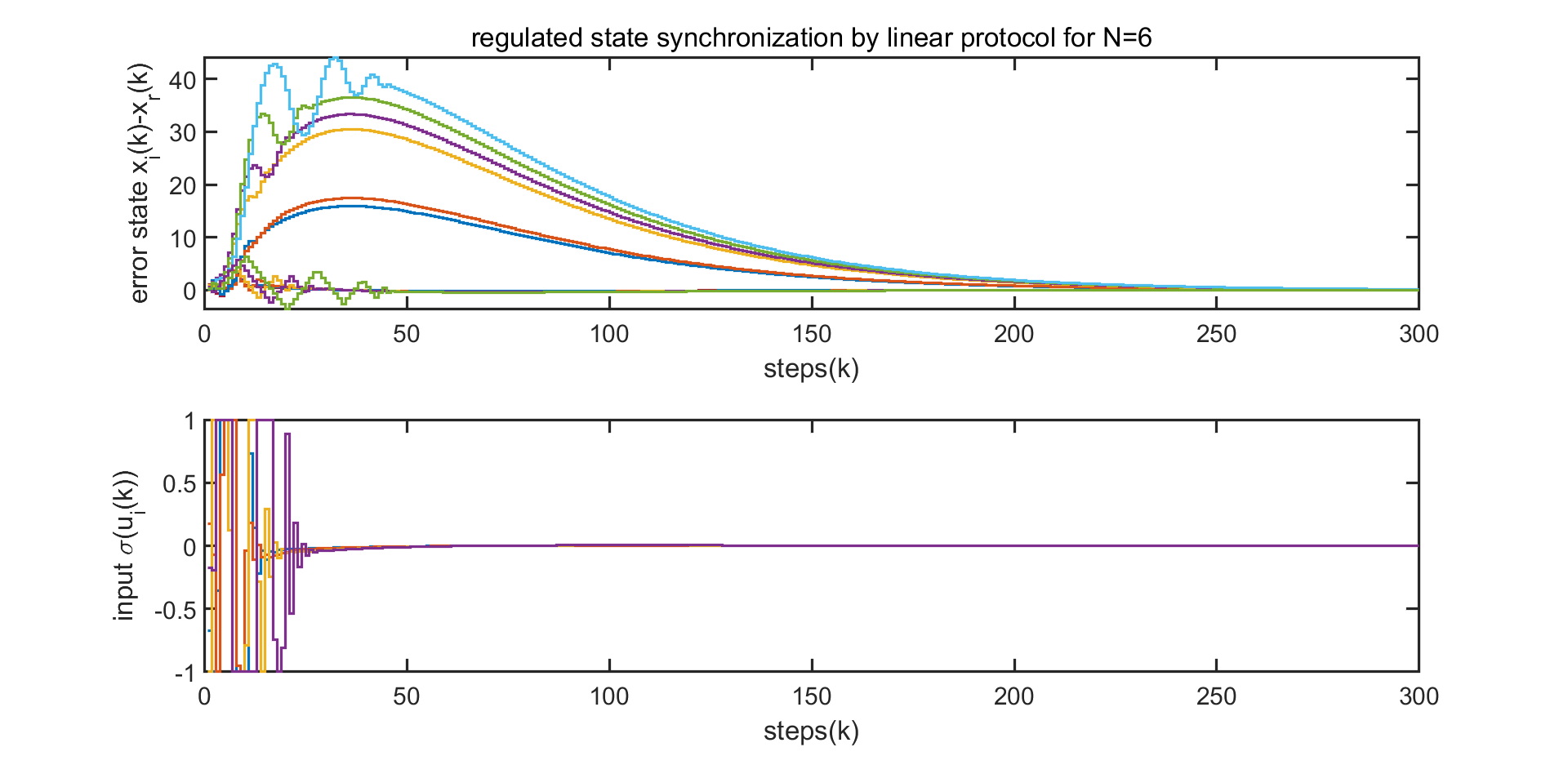} \centering
		\caption{Regulated state synchronization for
			MAS with communication graph
			II when $k_1=1.5$ and $k_2=2.5$.}\label{results_case2.3}
	\end{figure}
	
	\subsection*{\bf Case II: 6-agent graph}
	We consider a MAS with $6$ agents ($N=6$). The associated adjacency matrix to the communication network is assumed to be $\mathcal{A}_{II}$ where $a_{21}=a_{32}=a_{13}=a_{43}=a_{36}=a_{54}=a_{65}=1$. 
	
	In this case, we use protocol \eqref{protoclsim} and the corresponding simulation results are demonstrated in Figure~\ref{results_case2.1}-\ref{results_case2.3}.

	\begin{figure}[h!]
		\includegraphics[width=8cm]{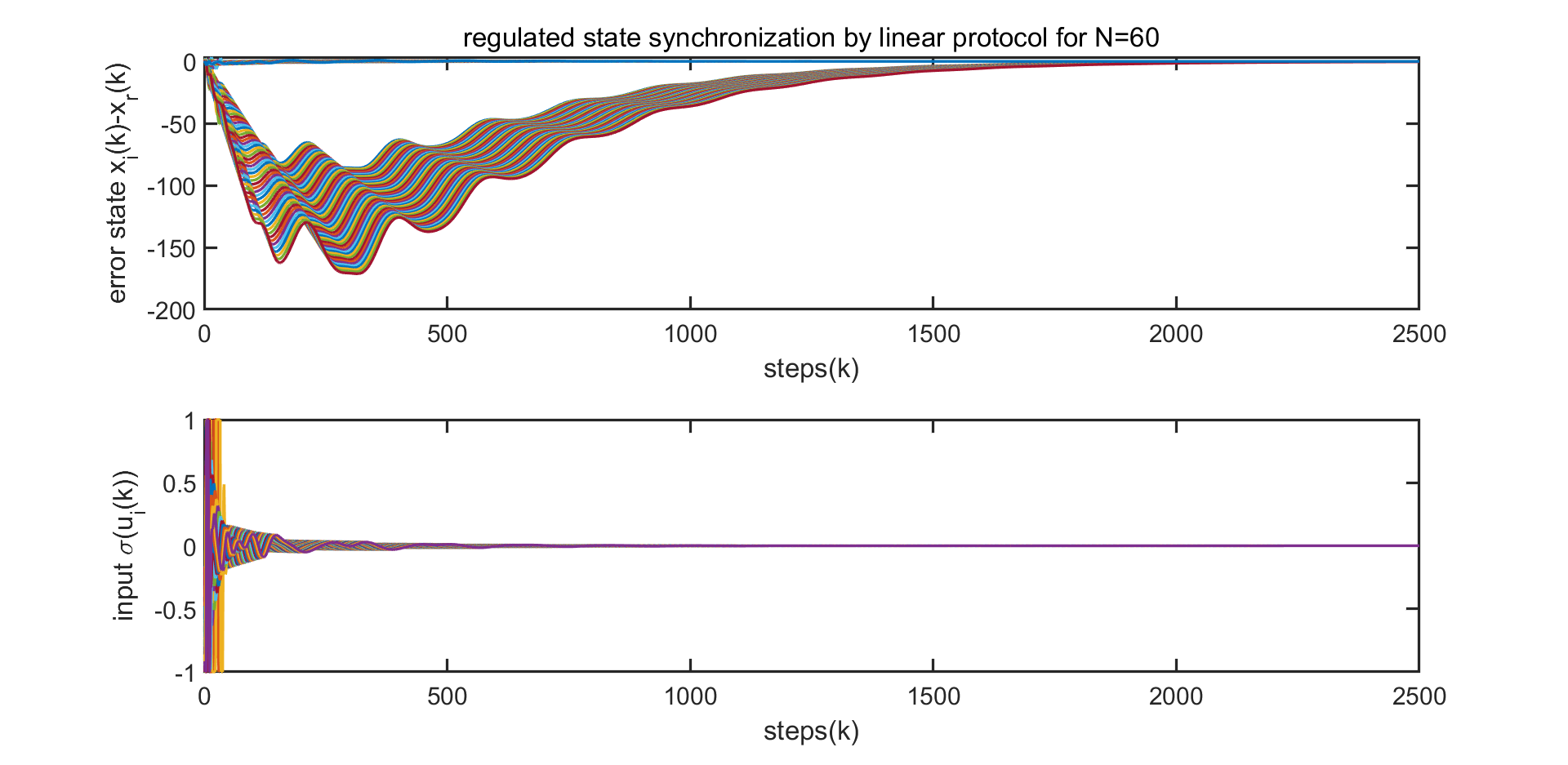} \centering
		\caption{{Regulated state synchronization for
				MAS with communication graph
				III when $k_1=0.5,k_2=1$.}}\label{results_case3.1}
	\end{figure}
	\begin{figure}[h!]
		\includegraphics[width=8cm]{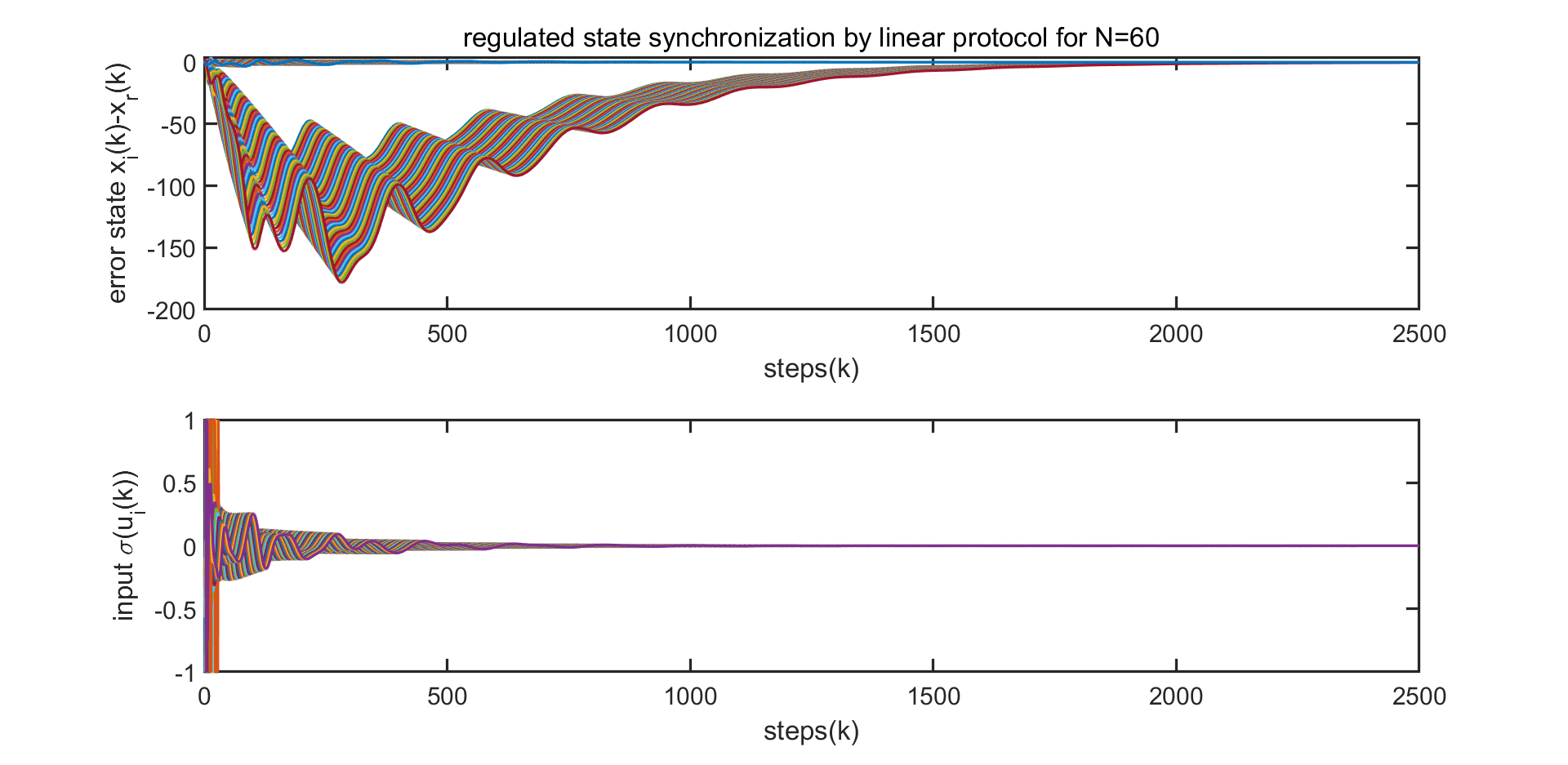} \centering
		\caption{Regulated state synchronization for
			MAS with communication graph
			III when $k_1=1$ and $k_2=2$.}\label{results_case3.2}
	\end{figure}
	\begin{figure}[h!]
		\includegraphics[width=8cm]{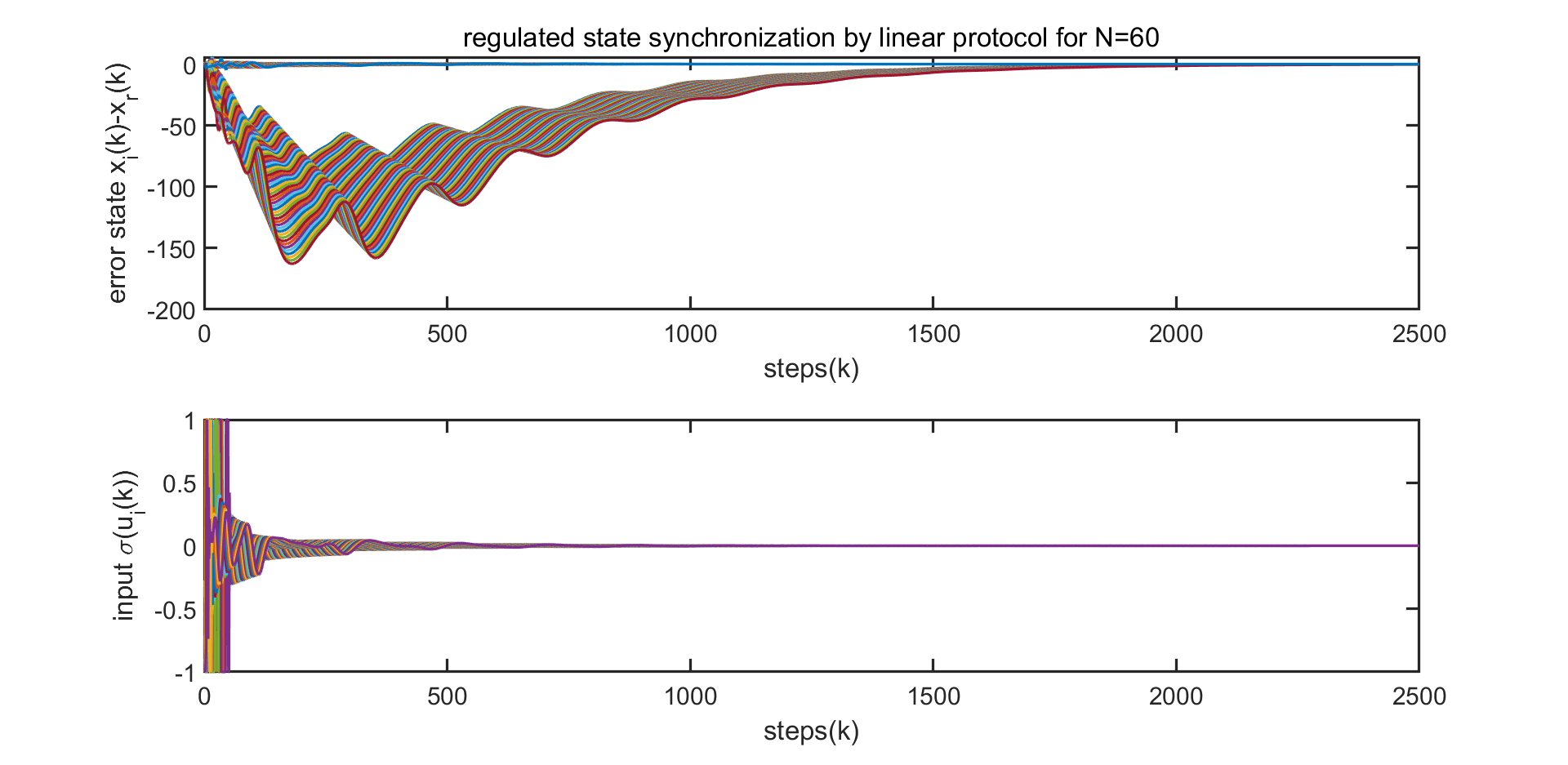} \centering
		\caption{Regulated state synchronization for
			MAS with communication graph
			III when $k_1=1.5$ and $k_2=2.5$.}\label{results_case3.3}
	\end{figure}
	
	\subsection*{\bf Case III: 60-agent graph}
	Then, we consider a MAS with $60$ agents (i.e. $N=60$) 
	and a directed loop graph, where the associated adjacency matrix is assumed to be $\mathcal{A}_{III}$ only with $a_{i+1,i}=a_{1,60}=1$ and $i=1,\cdots,59$.
	
	The simulation results by using protocol \eqref{protoclsim} are demonstrated in Figure
	\ref{results_case3.1}-\ref{results_case3.3}.

	All above simulation results show that
	the protocol with $k_1,k_2$ satisfying \eqref{cond1} can achieve synchronization. Meanwhile, 
	they also show that the
	protocol design is independent of the communication graph and is scale
	free so that we can achieve synchronization with one-shot protocol
	design, for any graph with any number of agents.
	
	\bibliographystyle{plain}
	\bibliography{referenc}
	
\end{document}